\documentclass[fleqn,12pt,twoside]{article}
\usepackage{espcrc1%
,latexsym%
,epsfig%
,color%
}

\usepackage{graphicx}
\usepackage[figuresright]{rotating}

\newcommand{\AmS}{{\protect\the\textfont2
  A\kern-.1667em\lower.5ex\hbox{M}\kern-.125emS}}

\title{Structure of baryons in a relativistic quark model}

\author{
Bernard Metsch\address[HISKP]{
Helmholtz-Institut f\"ur Strahlen- und Kernphysik (Theorie), \\
Universit\"at Bonn, Nu{\ss}allee 14-16, D-53115 Bonn, Germany}%
}
       
\begin{document}

\maketitle

\begin{abstract}
Baryonic excitation spectra, electroweak and strong decay properties
are discussed within a relativistically covariant constituent quark
model based on the instantaneous approximation to the three-body
Bethe-Salpeter equation.
\end{abstract}

\section{Introduction}
Baryonic resonance spectra exhibit some striking features:
Linear Regge-tra\-jec\-tor\-ies, which hint at a linear confinement
potential; moderately large hyperfine splittings (\textit{e.g.} the
$N-\Delta$-splitting) hinting at a strong spin-spin interaction;
parity doublets, such as \textit{e.g.} $N^*_{\frac{5}{2}^+}(1680)$-
$N^*_{\frac{5}{2}^-}(1675)$, which all are a challenge to explain
theoretically. The most successful approaches to account for these
have been constituent quark models (in non-relativistic or
'relativized' versions), see \textit{e.g.} the excellent review by
Capstick and Roberts~\cite{metsch_ref:1} and references therein, which
use one-gluon-exchange or Goldstone-boson-exchange as quark
interaction in addition to a linear confinement potential. Although
the results from such calculations are in general satisfactory, they
do not reproduce the details of the $N$-Regge-trajectory nor explain
the parity doublets found. Moreover, the role of the spin-orbit parts
of the residual interactions remains obscure. On top of this the
conventional constituent quark models have no real field theoretical
basis and lack relativistic covariance. As an extension of an earlier
relativistic quark model description of mesons~\cite{metsch_ref:2}, we
therefore developed a relativistic quark-model for baryons on the
basis of the three-particle Bethe-Salpeter equation.

The details of our approach are extensively described
in~\cite{metsch_ref:3}, here we shall merely quote the basic
assumptions and features. Starting point is the
Bethe-Salpeter-Equation for bound states of three fermions, which is a
homogeneous integral equation involving full quark propagators and
irreducible interaction kernels in terms of the 8 relative momentum
variables of the quarks.  In order to solve this equation we assumed,
inspired by the apparent success of the non-relativistic constituent
quark model, that the self-energy in the quark propagators can be
suitably approximated by introducing an effective, constituent quark
mass in the free Feynman-propagator and, furthermore, that the
interaction kernels do not depend on the relative energy variables of
the quarks in the rest-frame of the baryon. Although this also implies
a technical simplification (Salpeter equation), the main reason is
that we want to implement confinement by an instantaneous linearly
rising three-body potential. These assumptions, after introducing an
effective instantaneous kernel which approximates retardation effects
in two-body interactions, allow for a formulation of the resulting
Salpeter equation as an eigenvalue equation, which is solved by
expanding the amplitudes in a suitable large, but finite, basis.

\section{The model} 

Confinement is implemented by a string-like three body potential,
which rises linearly with inter-quark distances and comprises a spin
structure which was chosen such, that spin-orbit splittings are
suppressed, see~\cite{metsch_ref:4} for details. In order to account
for the hyperfine structure we adopted an effective two-body
interaction based on instanton effects, which has the decisive
property to solve the $U_A(1)$-problem in the pseudoscalar meson
spectrum~\cite{metsch_ref:2}. For two quarks it is a short-range
two-body interaction acting on quark pairs with vanishing spin
which are antisymmetric in flavour. Consequently this force does not
act on the flavour symmetric $\Delta$-resonances. The Regge-trajectory
in this sector is then used to determine the (non-strange) constituent
quark mass and the constant and string tension parameters of the
confinement potential. The three parameters of the instanton force are
determined from the ground-state octet-decuplet splittings. The
remaining spectrum is then a genuine prediction.

\section{Mass spectra}
\begin{figure}[htb]
\includegraphics[angle=270,width=1.0\textwidth]{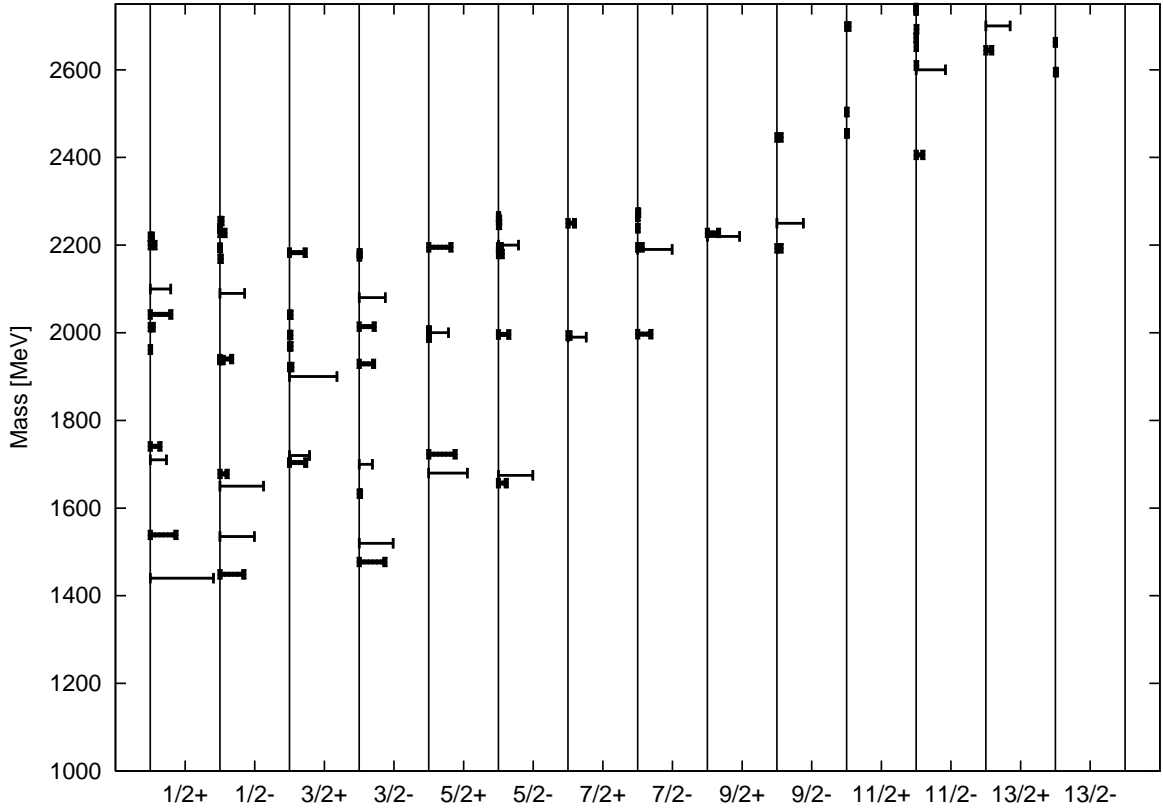}
\caption{
Decay amplitudes (proprtional to the square root of the partial decay
width) of strong $N^* \to N\pi$ decays. In each column
(\textit{i.e.} for each spin and parity $J\pi$) the
experimental value (thin horizontal bars at the experimaental resonance
position) is compared to the calculated
value (thick horizontal bars at the calculated resonance position).
}
\label{metsch_fig:1}
\end{figure}

The resulting mass spectra, see \textit{e.g.} Fig.~\ref{metsch_fig:1} 
for non-strange and strange baryons can be found
in~\cite{metsch_ref:4} and~\cite{metsch_ref:5}, respectively. In
general a very satisfactory description of the masses of states up to
2.5 GeV is found. The most prominent features are: Once the strengths
of the instanton induced interaction are fixed from the ground state
splittings the other prominent hyperfine splittings in the spectrum
can be explained quite naturally: In contrast to earlier
non-relativistic quark model calculations with the same interaction,
in the present relativistic setup the instanton-induced interaction is
strong enough to account almost quantitatively for the low position of
the Roper-resonances and its analogues in the strange sectors. In
addition, the $N$- and $\Lambda$- Regge trajectory can be very nicely
reproduced, indicating that this force leads to a constant shift in
$M^2$ for these states, in accordance with experimental data. Moreover
this force also accounts for the occurrence of the parity doublets
mentioned above: selectively those states from a particular shell (in
the harmonic oscillator classification), which show scalar diquark
configurations are lowered enough to become degenerate with some
states of the lower oscillator shell with opposite
parity~\cite{metsch_ref:4}.

\section{Electromagnetic observables}
On the basis of the Salpeter amplitudes, the baryonic vertex
function(amputated Bethe-Salpeter amplitude) can be reconstructed for
any on-shell baryon momentum und thus form factors and various
couplings can be calculated covariantly within the Mandelstam formalism, in impulse
approximation without any additional parameters. The details and
results of the calculation of static moments, electroweak form factors
and photon couplings of non-strange baryons have been published
recently~\cite{metsch_ref:12}. Here we will briefly discuss a novel
approach to calculate magnetic moments directly from the Salpeter
amplitudes and some new results for semi-leptonic and strong decays.

\subsection{Magnetic moments}
The interaction energy of a baryon in an external electromagnetic
field is given by the matrix element of $\hat j_\mu A^\mu$ where $\hat
j$ is the electromagnetic current. The magnetic moment is then the
coefficient of the term linear to a static magnetic field and can be
expressed as
\begin{equation}
\mu = -\frac{1}{4M}\varepsilon_{3jk} i 
\left[
\frac{\partial}{\partial \bar P^j}
\left\langle \bar M, S, S_3=S \right|
j^k(x^0,\vec 0)\left|\bar P, S, S_3=S \right\rangle
\right]_{\bar P = 0}\,,
\end{equation}
where $\left|\bar M, S\right\rangle$ and $\left|\bar P,
S\right\rangle$ involve the vertex functions of a baryon with mass $M$
and spin $S$ in the rest frame and of a baryon with momentum $\bar P$,
respectively.  Evaluating this expression we then find, that in
impulse approximation the magnetic moment can be expressed in momentum
space as the expectation value of a \textit{local} operator $\hat\mu$
with respect to the Salpeter amplitudes $\Phi^\Lambda_M$, normalized
to $\langle \Phi^\Lambda_M | \Phi^\Lambda_M \rangle = 2M$:
\begin{equation}
\mu =
\frac{\left\langle\Phi^\Lambda_M\right|\hat\mu\left|\Phi^\Lambda_M\right\rangle}{2M}
\,,\textrm{ where } 
\hat\mu^3 = \frac{\omega_1+\omega_2+\omega_3}{M}
\left(
\sum_{\alpha=1}^{3}\frac{\hat
e_\alpha}{2\omega_\alpha}\left(\hat \ell^3_\alpha+\hat
\Sigma^3_\alpha\right)
\right)
- \hat\mu^3_{C}\,,
\end{equation}
where $\omega_\alpha = \sqrt{m_\alpha^2+p_\alpha^2}$ represents the
relativistic energy of quark $\alpha$ with mass $m_\alpha$ and charge 
$\hat e_\alpha$, $\hat\ell$ and $\hat \Sigma$ represent the single particle angular
momentum-- and (twice the) spin--operator, respectively, and where 
$
\hat\mu_{C}^i = \varepsilon_{ijk}\frac{1}{M}
\sum_{\alpha_0}^3\frac{\hat
e_\alpha}{2\omega_\alpha}p_\alpha^k \sum_{\beta=1}^3\omega_\beta
\frac{\partial}{\partial p_\beta^j}\,.
$
corrects for the relativistic center of charge motion. We think that
this is a remarkable expression, to our knowledge not to be found in
the literature. The non-relativistic limit of this expression is of
course obvious. The results for octet baryons are given in Table~\ref{metsch_table:2}\,. 
\begin{table}[htb]
\caption{Magnetic moments of octet baryons in units of $\mu_N$.}
\label{metsch_table:2}
\newcommand{\m}{\hphantom{$-$}}
\newcommand{\cc}[1]{\multicolumn{1}{c}{#1}}
\renewcommand{\tabcolsep}{1.5pc} 
\renewcommand{\arraystretch}{1.2} 
\begin{tabular}{llllll}
\hline
{B}		& \cc{Exp.}		& \cc{Calc.}    & {B}		& \cc{Exp.}      	& \cc{Calc.}\\
\hline
${p}$		& \m{$2.792$}		& \m{$2.77$}	& $\Lambda$	& $-0.613\pm0.004$ 	& $-0.61$\\
${n}$   	& $-1.913$ 		& $-1.71$	& $\Sigma^+$	& \m{$2.458\pm 0.010$} 	& \m{$2.51$}\\
$\Xi^0$ 	& $-1.250\pm 0.014$	& $-1.33$ 	& $\Sigma^0$ 	& \m{($0.649$)} 	& \m{$0.75$}\\
$\Xi^-$		& $-0.65\pm0.08$	& $-0.56$  	& $\Sigma^-$	& $-1.160\pm 0.125$ 	& $-1.02$\\
\hline
\end{tabular}\\[2pt]
The experimental values, except for the value for $\mu_{\Sigma_0}$ which is taken via isospin invariance from
 $\mu_{\Sigma^+}$ and $\mu_{\Sigma^-}$, are given in ref. \cite{metsch_ref:9}.
\end{table}

\subsection{Semi-leptonic decays}

Some new, representative results for semi-leptonic decays, calculated from
the weak baryonic currents in the Mandelstam formalism, are listed in
Table~\ref{metsch_table:1}. 
\begin{table}[htb]
\caption{Decay rates and axial vector couplings of semi-leptonic decays of baryons.}
\label{metsch_table:1}
\newcommand{\m}{\hphantom{$-$}}
\newcommand{\cc}[1]{\multicolumn{1}{c}{#1}}
\renewcommand{\tabcolsep}{1.5pc} 
\renewcommand{\arraystretch}{1.2} 
\begin{tabular}{l@{ }c@{ }lllll}
\hline
&&&\multicolumn{2}{c}{$\Gamma$ $[10^{6}\textrm{s}^{-1}]$}&\multicolumn{2}{c}{$g_A/g_V$}\\
\hline
\multicolumn{3}{c}{Decay}	& 	\cc{Exp.}	& \cc{Calc.}    & \cc{Exp.}      & \cc{Calc.} \\
\hline
$n$ 	   &$\to$& $p\,e^-\bar\nu_e$   	        &                       &           &\m{$1.2670 \pm 0.0035$}    &\m{$1.21$}\\
$\Lambda$  &$\to$& $ p\,e^-\bar\nu_e$		&\cc{$3.16 \pm 0.06$}   &\cc{$3.10$}&{$-0.718 \pm 0.015$}       &{$-0.82$}\\
$\Sigma^+$ &$\to$& $\Lambda \,e^+\nu_e$		&\cc{$0.25 \pm 0.06$}	&\cc{$0.20$}& & \\
$\Sigma^-$ &$\to$& $\Lambda \,e^-\bar\nu_e$	&\cc{$0.38 \pm 0.02$}  	&\cc{$0.34$}& & \\
$\Sigma^-$ &$\to$& $ n\,e^-\bar\nu_e$		&\cc{$6.9 \pm 0.2$}    	&\cc{$4.91$}&\m{$0.340 \pm 0.017$}      &\m{$0.25$}	\\
$\Xi^0$    &$\to$& $\Sigma^+\,e^-\bar\nu_e$	&\cc{$0.93\pm 0.14$}   	&\cc{$0.91$}&\m{$1.32 {+0.21 \atop -0.17} \pm 0.05$} &\m{$1.38$}\\
$\Xi^-$    &$\to$& $\Sigma^0\,e^-\bar\nu_e$	&\cc{$0.5 \pm 0.1$}    	&\cc{$0.51$}& &	\\
$\Xi^-$    &$\to$& $\Lambda \,e^-\bar\nu_e$	&\cc{$3.3 \pm 0.2$}    	&\cc{$2.30$}&{$-0.25 \pm 0.05$}         &{$-0.27$}\\
$\Omega^-$ &$\to$& $ \Xi^0\,e^-\bar\nu_{e}$	&\cc{$68 \pm 34$}      	&\cc{$46$}  & &	\\
$\Lambda$  &$\to$& $ p\,\mu^-\bar\nu_{\mu}$	&\cc{$0.60 \pm 0.13$}  	&\cc{$0.47$}& &	\\
$\Sigma^-$ &$\to$& $ n\,\mu^-\bar\nu_{\mu}$	&\cc{$3.04 \pm 0.27$}  	&\cc{$1.60$}& &	\\
$\Xi^-$    &$\to$& $\Lambda\,\mu^-\bar\nu_{\mu}$&\cc{$2.1 \pm 1.3$}   	&\cc{$1.04$}& &	\\
\hline
\end{tabular}\\[2pt]
The experimental values are given in ref. \cite{metsch_ref:9}.
\end{table}

\section{Strong decays and the problem of ''missing resonances''}
\begin{minipage}[b]{0.3\textwidth}
\begin{center}
\includegraphics[width=1.0\textwidth]{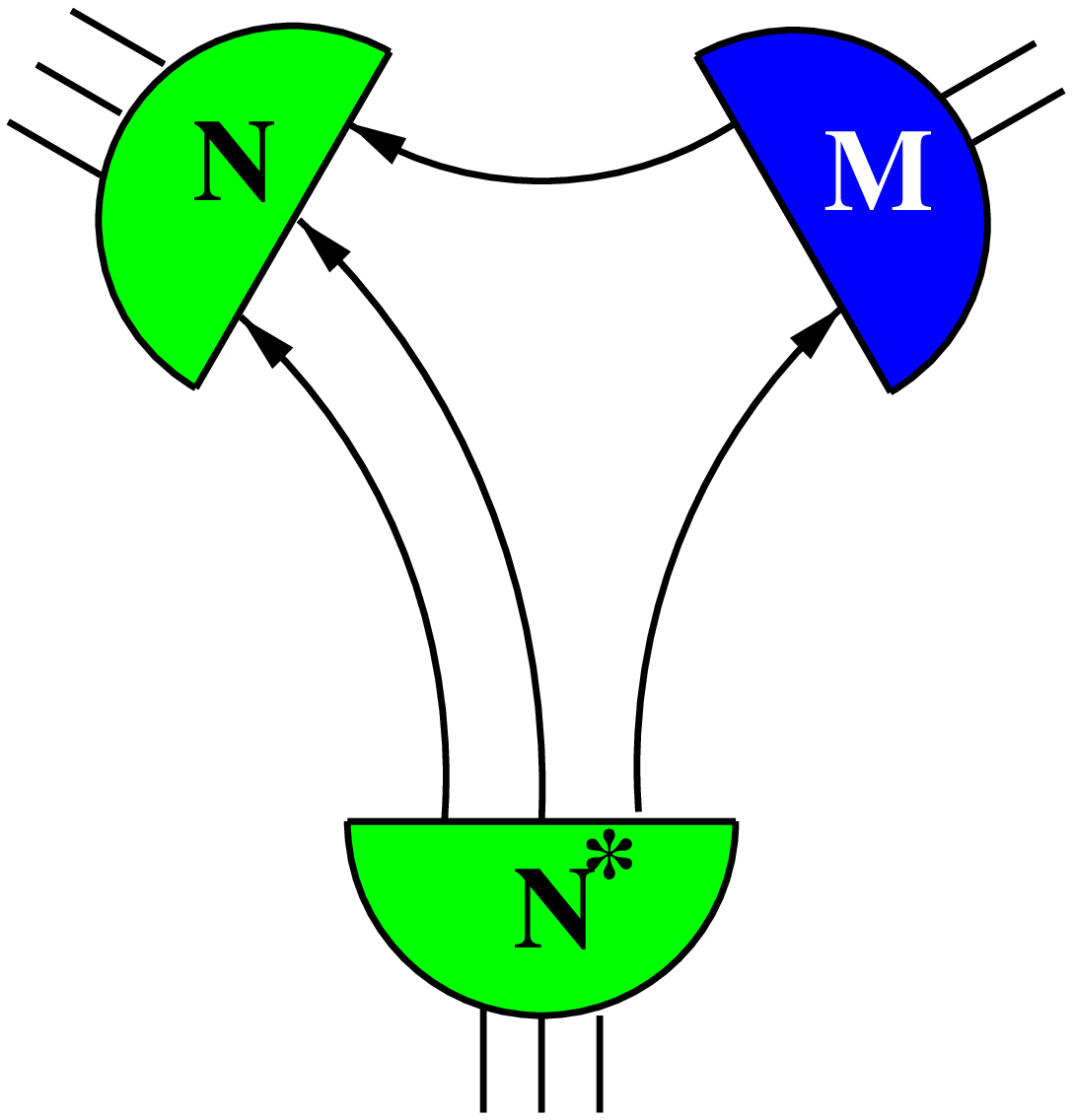}\\
\label{metsch_fig:2} 
\end{center}
Figure 2. 
{Lowest order quark-loop contribution to the strong 
$N^*\to N+M$-decay of excited baryons.}
\end{minipage}
\hspace{\fill}
\begin{minipage}[b]{0.6\textwidth}
In the framework of the Mandelstam formalism the amplitude for 
the strong mesonic decay of excited baryons can be obtained in lowest
order by evaluating the simple quark loop diagram of
Fig.~{2}, 
which involves the vertex functions (amputated
Bethe-Salpeter-amplitudes) of the participating meson, obtained from a
previous calculation on mesons~\cite{metsch_ref:2}, and
baryons. Although in general the calculated partial widths are too
small to account for the experimental values quantitatively,
appreciable decay widths are found only for the well established
resonances, the predicted values for higher lying resonances being in general
smaller by at least an order of magnitude, see also
Fig.~\ref{metsch_fig:1}, thus explaining why these have not been
observed so far in elastic pion-nucleon scattering. This observation
is in accordance with the findings cited in~\cite{metsch_ref:1}\,.
\end{minipage}

\section{Conclusion}

In conclusion we think that we have demonstrated, that a fully
relativistic treatment of the quark dynamics and their electromagnetic
couplings on the basis of the instantaneous Bethe-Salpeter equation with 
quark forces from a confinement potential and an instanton-induced
interaction indeed leads to an encouraging description of the
major features of the baryonic excitation spectrum upto 3 GeV, including numerous electromagnetic and strong decay observables, 
This description still can and, 
in view of the approximate treatment of retardation effects from the
two-body interaction, must be improved.

Major contributions by Christian Haupt, Dirk Merten, Sascha Migura
and Herbert Petry are gratefully acknowledged. We thank the DFG for
financial support.

\end{document}